%% file: Tuchming_Higgs_JP_D0_2013.tex
\documentclass[epj,twocolumn,english]{webofc}

\usepackage[varg]{txfonts}   
\input{bobo_macros.tex}

\renewcommand{\dzero}{D0}
\woctitle{LHCP 2013}

\begin{document}
\selectlanguage{english} 

\title{Spin/parity of Higgs-like particle at D0}

%
%

\author{Boris Tuchming\inst{1}
\fnsep\thanks{\email{tuchming@cea.fr
    }} for the \dzero\ Collaboration.
}

\institute{
CEA Saclay -  Irfu/SPP - France.
         }

\def\whl{\ensuremath{WH\to \ell\nu b\bar b}}
\def\zhl{\ensuremath{ZH\to \ell^+\ell^- b\bar b}}
\def\zhv{\ensuremath{ZH\to \nu\bar\nu b\bar b}}
\abstract{%
We present prospects for tests of different spin and parity hypotheses for a particle $H$ of mass 125~GeV produced in association with a vector boson and decaying into a pair of b-quarks. We use the combined analysis of  the $WH\to \ell\nu b\bar b$,
 $ZH\to \ell^+\ell^- b\bar b$, and 
 $ZH\to \nu\bar\nu b\bar b$  
channels based on the full Run~II dataset 
collected  at  $\sqrt s=1.96$~TeV with the \dzero\ detector at the Fermilab Tevatron collider.
}

\maketitle
\section{Introduction }

\label{intro}

In the standard model (SM), the electroweak symmetry breaking mechanism results from 
the existence of a single elementary scalar field doublet
that acquires a non-zero vacuum expectation value
and manifests itself as a scalar particle, the Higgs boson, the mass of which is free parameter in the model.

Finding the  Higgs boson has been one of the most topical goals
of particle physicists in the last decades.
In summer 2012, 
the CDF and D0 Collaborations  reported excesses
above background expectations in the $H\rightarrow b\bar b$
search channels~\cite{bib:CDFhbb,bib:D0hbb}. 
Their combination yields an excess at the three s.d.\ level,
consistent with the production of a Higgs boson of mass $M_H \approx 125$~\egev~\cite{bib:TeVhbb}.
At the same time, the ATLAS and CMS Collaborations
reported the discovery of a new particle 
at the five standard deviation (s.d.) level,
consistent with the observation of a Higgs boson of $M_H \approx 125$~\egev\
in the $H\to ZZ$ and $H \to \gamma\gamma$ channels~\cite{bib:atlas-hdiscovery,bib:cms-hdiscovery}.

After the discovery of this new particle, comes the time to establish
its properties, such as mass, spin, parity, and couplings strengths, and check if it is the Higgs boson of the SM.
Testing properties in  the Tevatron $VH\to V b\bar b$ modes would provide
a consistent complementary information relative to  the numerous spin/parity analyses performed at the LHC in the $H\to\gamma\gamma$, $H\to ZZ$, and $H\to W^+W^-$ modes.
This proceedings discusses tests of spin/parity for the new particle of 125~GeV using the  SM Higgs boson
$VH\to V b\bar b$ search channels.
The analysis is based on the full Run~II dataset
consisting of $\sim 10~\fbinv$ of $p\bar p$ collision
recorded  at  $\sqrt s=1.96$~TeV with the \dzero\ detector at Fermilab.

\section{Principles}

In general spin/parity of a particle affects angular correlations of its decay products,
but also excitation curve behavior near production threshold.
For example, the cross-section for the process $\epem\to Z^* \to Z H$ at center-of-mass energy $\sqrt s$ exhibits a behavior
varying like $\sigma \sim \beta=\sqrt{ \frac{s -(M_H+M_Z)^2 }
{s -(M_H-M_Z)^2}}$
due to $s$-wave contributions if $J^{P}(H)=0^+$, as for the SM Higgs boson~\cite{Choi:2002jk}.
This is modified into $\sigma\sim \beta^3$ (due to $p$-wave) if $J^{P}=0^-$. For particle with  $J^{P}(H)=2^+$, many possibilities are allowed for the $HZZ$ coupling,
but certain models end up with the cross-section dominated by the $d$-wave terms, resulting in a $\sigma\sim \beta^5$  dependence.

At hadronic colliders, such as the Tevatron,
the same effect is expected in the $q\bar q \to VH$ process. However, 
the effective center of mass energy $\sqrt{\hat s}$  is not fixed
and depends on both  cross-section dynamics and the parton density functions.
Thus, the spin/parity of the particle $H$ affects the shape of the differential cross-section as a function of effective energy, $\sqrt{\hat s}$,  of the process  $p\bar p \to VH\to V b\bar b$~\cite{bib:Elis}.

\begin{figure}[!]
\centering
\includegraphics[width=0.45\textwidth,clip]{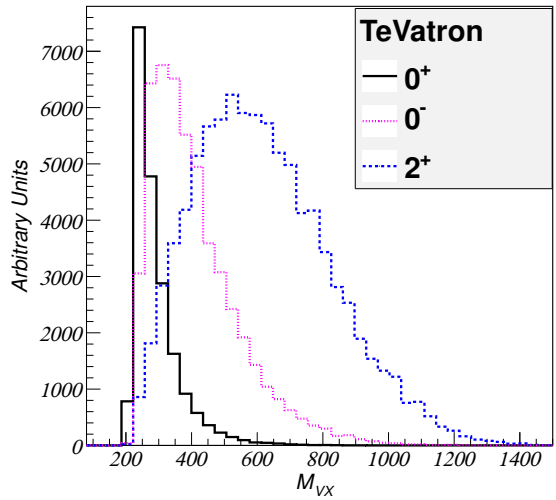}
\caption{Distribution of the total invariant mass of the $VH$  system at Tevatron for different spin/parity hypotheses (figure extracted from Ref.~\cite{bib:Elis}).}
\label{fig:theory_mtot}       
\end{figure}

This later property can be exploited in the $VH\to V b\bar b$ search modes by using discriminating variables related to the
total energy: either the overall mass of reconstructed objects, or their transverse mass for final states with missing transverse energy ($\etmis$) due to neutrinos.
The difference in shape for such observables is shown in Fig.~\ref{fig:theory_mtot}.
In the following we use the  $J^{P}=0^+$, $J^{P}=0^-$, and  $J^{P}(H)=2^+$ signal models as described in~\cite{bib:Elis}
for particles of mass 125~\gev.
The signal Monte Carlo samples are generated by the Madgraph 5 version 1.4.8.4 generator~\cite{Alwall:2011uj}.

\section{Data analysis}

The data analysis follows closely the steps of the search for SM Higgs boson,
in the \whl~\cite{dzWHl},
 \zhl~\cite{dzZHll1-jul12}, and 
 \zhv~\cite{dzZHv2} search channels,
except that in the final step, we employ  the overall mass or transverse mass of the selected events as final discriminant instead of a dedicated multivariate discriminant.
These analyses rely on
good $b$-tagging efficiency, good dijet mass resolution, high-\pt\ lepton acceptance,
good modeling of the \etmis, and good modeling of the $V$+jet background.
The tagging of $b$-jets is performed with a boosted decision tree (BDT) $b$-tagger.

The main feature of the selections are as follows:
\begin{itemize}
\item The \whl\ search channel relies
on the selection of isolated high-\pt\ electrons or muons,
at least two jets, and large \etmis.
A BDT
discriminant
is used to discriminate against multijet
background. 
The sample of selected events is
divided into exclusive subchannels according to lepton flavors
and $b$-tagger outputs: 1 tight tag, 2 loose tag, 2 medium tag, 2 tight tag.
To discriminate signal events from background events, a BDT final discriminant is constructed 
for each lepton flavor, jet multiplicity, and $b$-tagging category.  
In addition
to kinematic variables, the inputs to the 
final discriminants include the $b$-tagger output and the output from the multijet discriminant.
The BDT  is trained against the SM Higgs boson signal, but its discriminating power happens to be
close to optimal for $0^-$ or $2^+$ signal,
as can be seen in Fig.~\ref{fig:jp_signal_enhancement} (bottom), where the BDT output (denoted MVA) distribution
is shown for the 2-tight-tag sample.

\begin{figure}[!htb]
\centering
\includegraphics[width=0.40\textwidth,clip]{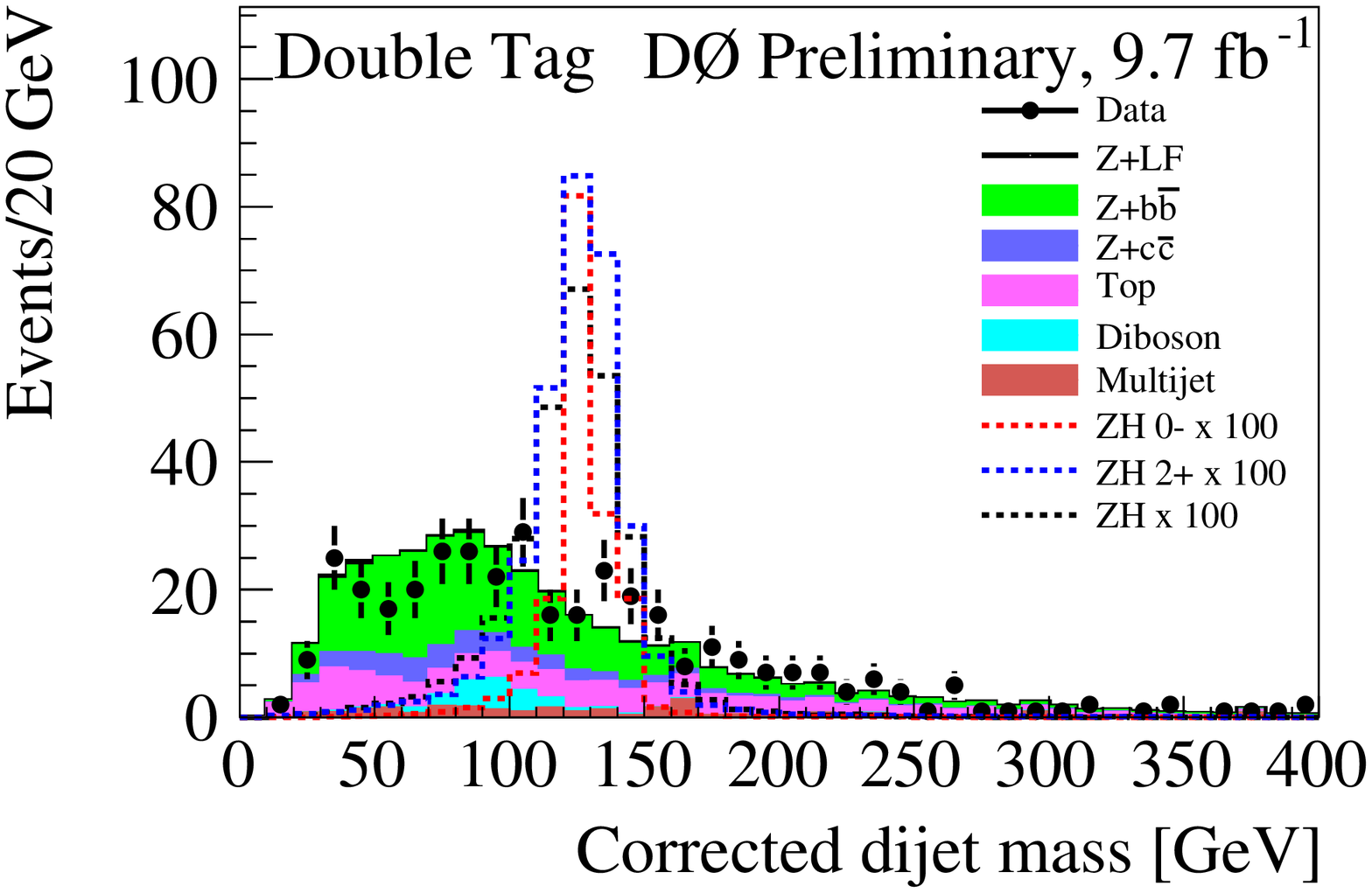}
\includegraphics[width=0.40\textwidth,clip]{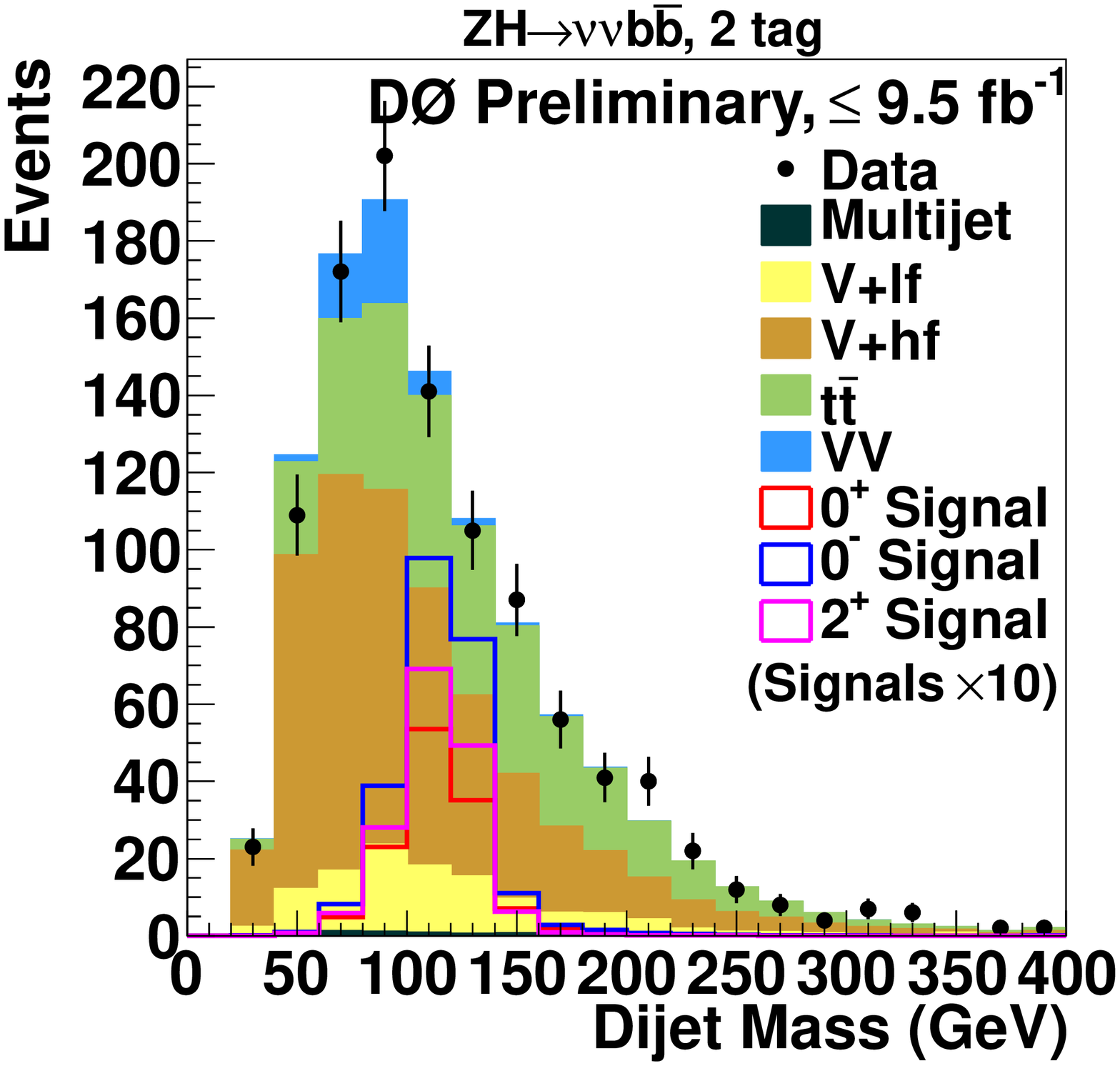}
\includegraphics[width=0.40\textwidth,clip]{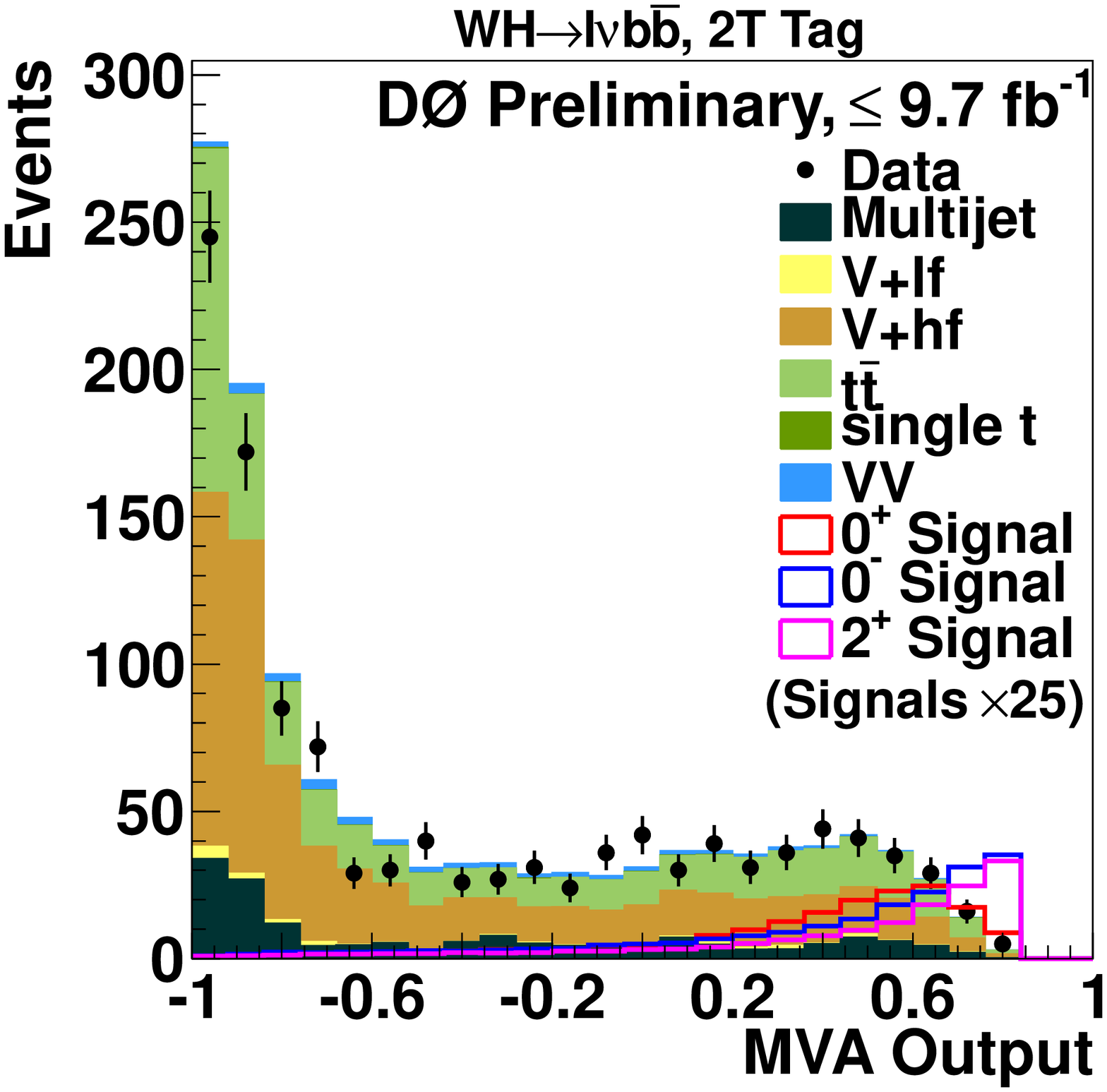}
\caption{Distributions of variables employed to enhance signal over background purity, for
 (top) the \zhl, (middle) \zhv, and (bottom) \whl\ channels.}
\label{fig:jp_signal_enhancement}
\end{figure}

\item
The \zhl\ analysis requires two isolated
charged leptons of opposite charge and at least two jets. 
The lepton acceptance is increased thanks to secondary channels
with loose lepton identification criteria:  electrons in the \dzero\ inter-cryostat region,
and isolated tracks not reconstructed in the muon spectrometer.
A kinematic fit corrects the 
measured jet energies to their best fit values according to the constraints
that the dilepton invariant mass should be consistent with the $Z$ boson mass
$M_Z$ and the total transverse momentum of the leptons and jets 
should be consistent with zero.
A first jet is demanded to pass tight b-tagging criteria.
The events are then divided into  double-tag and single-tag subchannels depending 
on whether a second jet passes a loose $b$-tagging requirement.
Figure~\ref{fig:jp_signal_enhancement} (top) presents the distribution of the dijet invariant mass
for selected events in the double-tag sample.

\item 

The \zhv\ analysis selects events with large \etmis\ and
two jets.
This search is also
sensitive to the $WH$ process when the charged lepton
from $W\to\ell\nu$ decay is not identified.
To reduce the multijet background a dedicated BDT discriminant is employed.
Events are split in two $b$-tagging subchannels using the sum of the 
$b$-tagging discriminant outputs of the two jets.
Figure~\ref{fig:jp_signal_enhancement} (middle) presents the distribution of the dijet invariant mass
for selected events in the double-tag sample.

\end{itemize}

\begin{figure}[!htb]
\centering
\includegraphics[width=0.40\textwidth,clip]{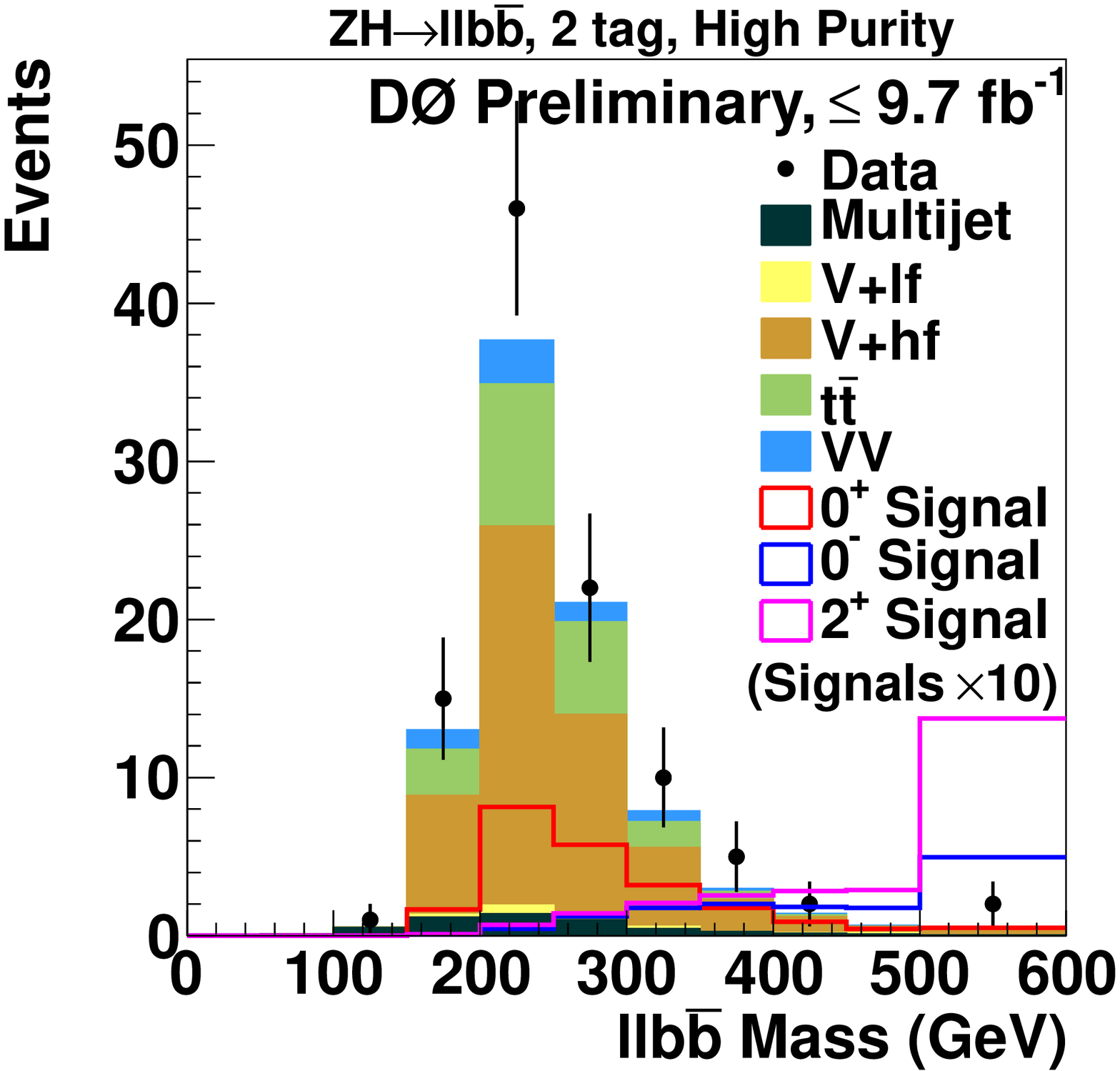}
\includegraphics[width=0.40\textwidth,clip]{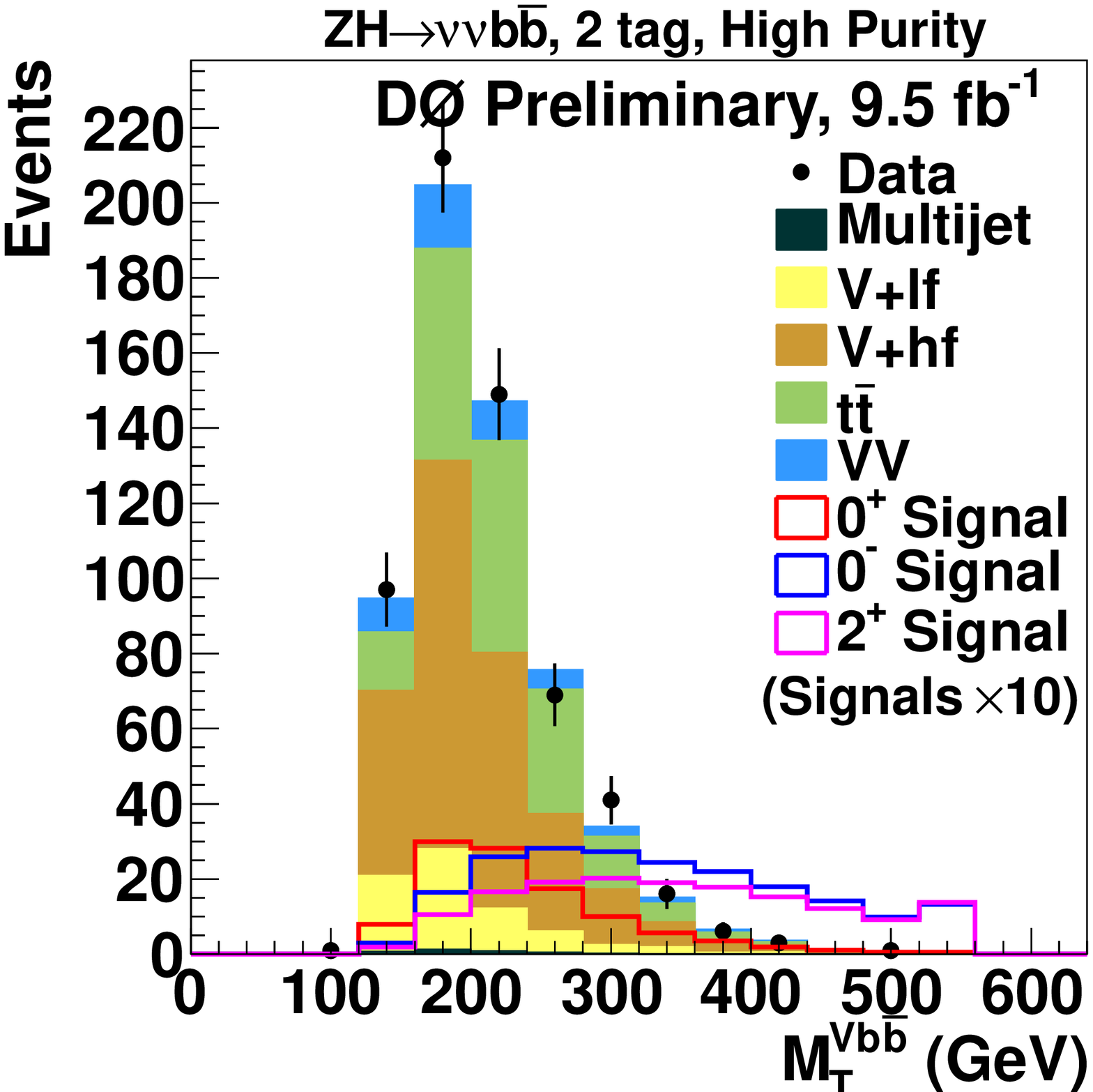}
\includegraphics[width=0.40\textwidth,clip]{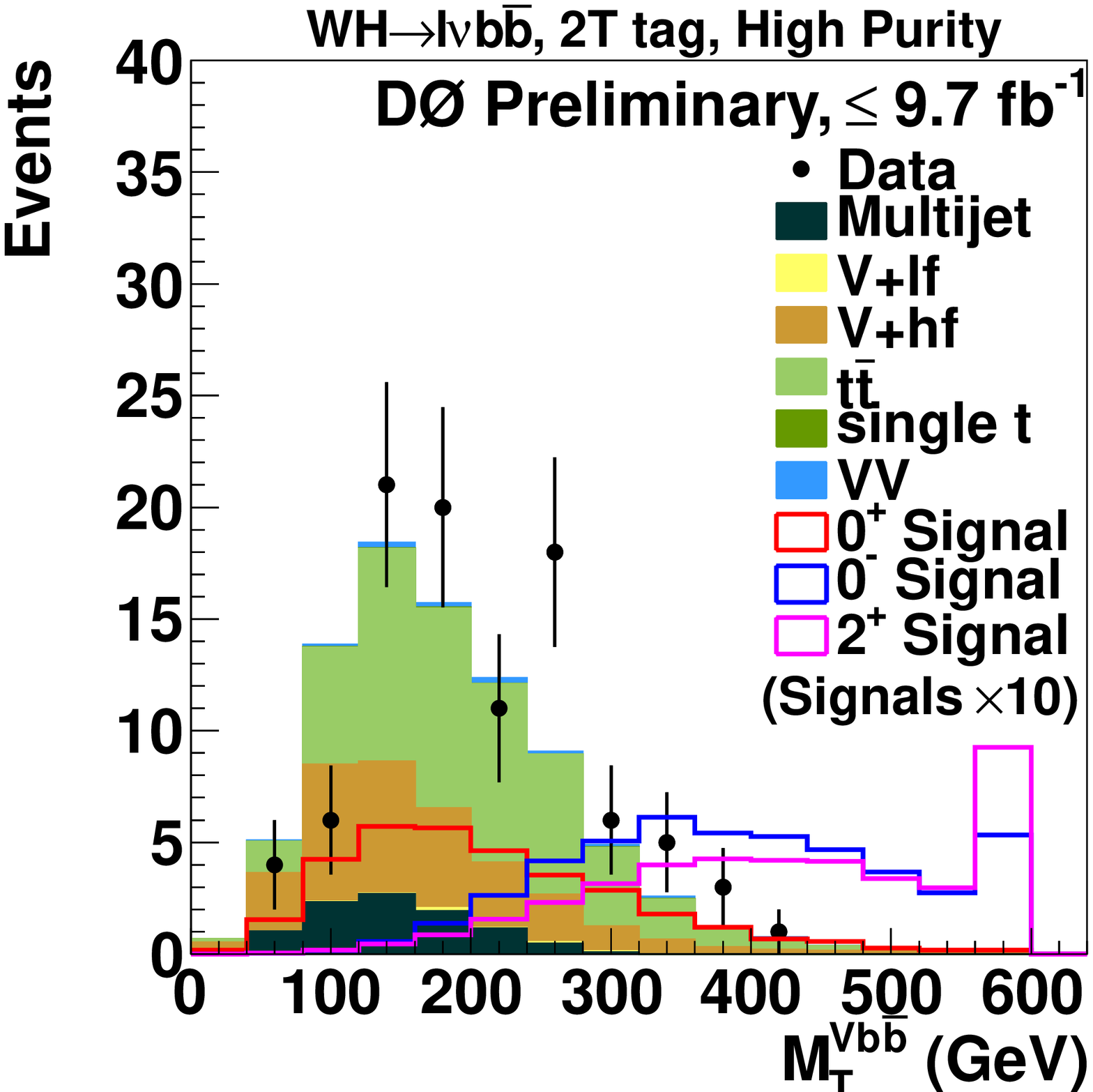}
\caption{Top, Overall mass for the \zhl\  high-purity sample at  the final selection stage. Middle and bottom, overall transverse mass  at the final selection stage
for the high purity samples of the \zhv\ and $\whl$ channels, respectively.}
\label{fig:jp_signal_discriminant}
\end{figure}

To enhance the signal over background ratio at the final selection stage, the selected samples are split into higher and lower-purity regions according to the dijet
invariant mass for \zhl\ and \zhv, and the final BDT discriminant  output for $\whl$.
The $\whl$ channel defines low-, medium-, and high-purity regions 
according to  $-1 <{\rm {MVA}} \le0$,  $0 <{\rm {MVA}} \le0.5$,  and $0.5< {\rm {MVA}} $, respectively.
In the $\zhv$ channel, the low purity region is defined by  $M_{jj}<70$~\gev\ or $M_{jj}>150$~\gev, and the high purity region by $70\le M_{jj} \le150~\gev$.
In the $\zhl$ channel, the low purity region is defined by  $M_{jj}<100$~\gev\ or $M_{jj}>160$~\gev, and the high purity region by $100 \le M_{jj}\le 160~\gev$.

As discussed above,
the main variables to discriminate the $2^+$ and $0^-$ signals against the background and the $0^+$ Higgs boson are the total dilepton+dijet mass in the $\ell\ell b\bar b$ final states, and the total transverse mass for the $\ell\nu b\bar b$
and  $\nu\bar\nu b\bar b$ final states. Their distributions are shown in Fig~.\ref{fig:jp_signal_discriminant}.

\section{Expected results}

The final step of the analysis consists in constructing a log-likelihood ratio test-statistic $LLR = -2 \ln( L_{H_1}/L_{H_0})$,
based on the final discriminant distributions of candidate events,
where $L_{H_0}$ is the likelihood function for the SM-Higgs-boson-plus-background
hypothesis ($0^+$-plus-background), and
$L_{H_1}$ is either the likelihood function for the $0^-$-plus-background hypothesis or the
$2^+$-plus-background hypothesis.
In this test we assume that cross-sections times branching fractions are identical to the SM one.
In the LLR calculation the signal and background rates
are functions of the systematic uncertainties which are taken into
account as nuisance parameters with Gaussian priors.
Their degrading effect is reduced by
fitting signal and background contributions to the data by 
maximizing the profile likelihood function for the 
$H_0$ and $H_1$ hypotheses separately, appropriately
taking into account all correlations between the systematic
uncertainties~\cite{bib:collie}.

\begin{figure}[!h]
\centering
\includegraphics[width=0.5\textwidth,clip]{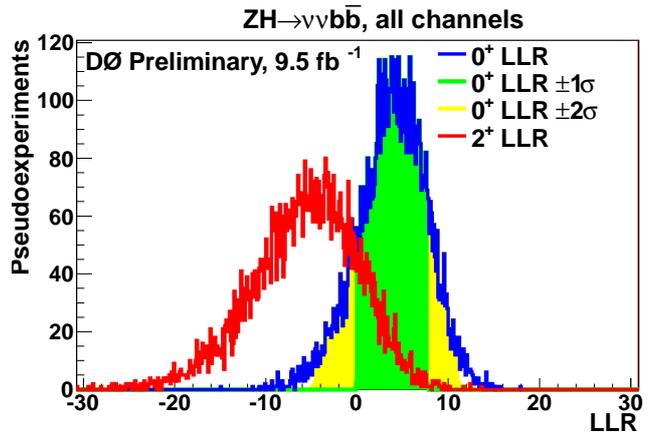}
\caption{
Distribution of $LLR$ in the \zhv\ channel for random $0^+$-plus-background and
$2^+$-plus-background pseudo-experiments.}
\label{fig:jp_final_pe_distribution}
\end{figure}

Figure~\ref{fig:jp_final_pe_distribution} shows the distributions of pseudo-experiments for the
$LLR$ constructed for $\zhv$ channel only, with $H_1$ being the  $2^+$-plus-background hypothesis. Two set of random pseudo-experiments
have been drawn, under the $H_1$ and $H_0$ hypotheses, respectively.
This figure demonstrate the good separation between the two hypotheses.
In the SM hypothesis, the expected $2^+$ confidence level,
$P(LLR \ge LLR^{obs} | H_1= 2^+{\mbox{-plus-background}})$ amounts to approximately 3.5\%.
Further separation is expected once the three channels will be combined.

\section{Conclusion}
The Tevatron $VH \to V b\bar b$ Higgs search channels can be used to probe
the spin/parity properties of the newly discovered Higgs-like particle.
The \dzero\ Collaboration has started an analysis based on these search channels to test the
$J^{P}=0^-$ and $J^{P}=2^+$ hypotheses. A good separation power between the $J^{P}=2^+$
and the SM Higgs boson ($J^{P}=0^+$) is demonstrated with
the   \zhv\ channel only. More separation is expected once all channels will be combined.
The results of this analysis are expected to be released soon.

\section{Acknowledgments}

We thank the staffs at Fermilab and collaborating institutions,
and acknowledge support from the
DOE and NSF (USA);
CEA and CNRS/IN2P3 (France);
MON, NRC KI and RFBR (Russia);
CNPq, FAPERJ, FAPESP and FUNDUNESP (Brazil);
DAE and DST (India);
Colciencias (Colombia);
CONACyT (Mexico);
NRF (Korea);
FOM (The Netherlands);
STFC and the Royal Society (United Kingdom);
MSMT and GACR (Czech Republic);
BMBF and DFG (Germany);
SFI (Ireland);
The Swedish Research Council (Sweden);
and
CAS and CNSF (China).
We also thank Ken Herner who helped in the preparation of this talk.
%
%
%

\end{document}

%% file: bobo_macros.tex
\newcommand{\invisible}[1]{}

\def \etmis {\mbox{\ensuremath{E\kern-0.6em\slash_T}}}
\def \etmiss {\mbox{\ensuremath{E\kern-0.6em\slash_T}}}

\newcommand{\rien}{}

\newcommand{\INVISIBLE}[1]{}

\newcommand{\dzero}  {\ensuremath{\mathrm{D\O}}}

\newcommand{\epem}{\ensuremath{\rien{e^+e^-}}}

\newcommand{\gev}  {\ensuremath{\mathrm{ GeV}}}

\newcommand{\egev}{\gev}

\newcommand{\fbinv}{\ensuremath{\mathrm{ fb^{-1}}}}

\newcommand{\pt}{{\ensuremath{ p_T}}}

\newcommand{\BEA}{\begin{eqnarray}}
\newcommand{\EEA}{\end{eqnarray}}

\def\beq  {\begin{equation}}
\def\eeq  {\end{equation}}

%% file: Tuchming_Higgs_JP_D0_2013.bbl
\begin{thebibliography}{}
%
%




\bibitem{bib:CDFhbb} 
  T.~Aaltonen {\it et al.}  [CDF Collaboration],
  Phys.\ Rev.\ Lett.\  {\bf 109}, 111802 (2012).

\bibitem{bib:D0hbb} 
  V.~M.~Abazov {\it et al.}  [D0 Collaboration],
  Phys.\ Rev.\ Lett.\  {\bf 109}, 121802 (2012).

\bibitem{bib:TeVhbb} 
  T.~Aaltonen {\it et al.}  [CDF and D0 Collaborations],
  Phys.\ Rev.\ Lett.\  {\bf 109}, 071804 (2012).

\bibitem{bib:atlas-hdiscovery}
  G.~Aad {\it et al.}  [ATLAS Collaboration],
  Phys.\ Lett.\ B {\bf 716}, 1 (2012).


\bibitem{bib:cms-hdiscovery}  
  S.~Chatrchyan {\it et al.}  [CMS Collaboration],
  Phys.\ Lett.\ B {\bf 716}, 30 (2012).


\bibitem{Choi:2002jk} 
  S.~Y.~Choi, D.~J.~Miller, 2, M.~M.~Muhlleitner and P.~M.~Zerwas,
  Phys.\ Lett.\ B {\bf 553}, 61 (2003)





\bibitem{bib:Elis}
  J.~Ellis, D.~S.~Hwang, V.~Sanz and T.~You,
  %
  JHEP {\bf 1211}, 134 (2012)

\bibitem{Alwall:2011uj} 
  J.~Alwall, M.~Herquet, F.~Maltoni, O.~Mattelaer and T.~Stelzer,
  JHEP {\bf 1106}, 128 (2011)


\bibitem{dzWHl} V.~M.~Abazov {\sl et al.} (D0 Collaboration), Phys.\ Rev.\ Lett {\bf 109}, 121804 (2012).
V.~M.~Abazov {\sl et al.} (D0 Collaboration), arXiv:1301.6122 (2013), accepted by Phys.\ Rev.\ D

\bibitem{dzZHll1-jul12}
V.~M.~Abazov {\sl et al.} (D0 Collaboration), Phys.\ Rev.\ Lett {\bf 109}, 121803 (2012).
V.~M.~Abazov {\sl et al.} (D0 Collaboration), arXiv:1303.3276 (2013), accepted by Phys.\ Rev.\ D.


\bibitem{dzZHv2}
V.~M.~Abazov {\sl et al.} (D0 Collaboration), Phys.\ Lett.\ B {\bf 716}, 285 (2012).





\bibitem{bib:collie}
W. Fisher, FERMILAB-TM-2386-E (2006).




\end{thebibliography}
